# Investigations of Graphene on SrTiO$_3$ Single-Crystal using Confocal Raman Spectroscopy

S. Shrestha[1], C. S. Chang[2,3], S. Lee[2], N. L. Kothalawala[4], D. Y. Kim[4], M. Minola[5], J. Kim[2], A. Seo[1]

[1]*Department of Physics and Astronomy, University of Kentucky, Lexington, KY 40506, USA*
[2]*Department of Mechanical Engineering, Massachusetts Institute of Technology, Cambridge, MA 02139, USA*
[3]*Research Laboratory of Electronics, Massachusetts Institute of Technology, Cambridge, MA 02139, USA*
[4]*Department of Chemistry, University of Kentucky, Lexington, KY 40506, USA*
[5]*Max-Planck-Institut für Festkörperforschung, D-70569 Stuttgart, Germany*

## Abstract

Graphene layers placed on SrTiO$_3$ single-crystal substrates were investigated using temperature-dependent confocal Raman spectroscopy. This approach successfully resolved distinct Raman modes of graphene that are often untraceable in conventional measurements due to the strong Raman scattering background of SrTiO$_3$. Information on defects and strain states was obtained for a few graphene/SrTiO$_3$ samples that were synthesized by different techniques. This confocal Raman spectroscopic approach can shed light on the investigation of not only this graphene/SrTiO$_3$ system but also various two-dimensional layered materials whose Raman modes interfere with their substrates.

Understanding the properties of graphene deposited on various substrates or templates is essential for the advancement of graphene-based device applications. The physical properties of the two-dimensional (2D) carbon atoms in a honeycomb lattice are strongly influenced by the substrate material[1-5]. For example, graphene layers on hexagonal boron nitrides can exhibit a few orders of magnitude higher mobilities than those on silicon substrates[3]. Graphene interfacing with high-$k$ dielectric materials is also promising because high electric fields can be applied directly, achieving ultra-high carrier densities or improved field-effect transistor effects in the graphene layer[5,6]. Strontium titanate ($SrTiO_3$) crystals are considered one of the candidate materials for this perspective due to their high dielectric permittivity at low temperatures[7]. Recently, graphene layers on $SrTiO_3$ single crystals (graphene/$SrTiO_3$) are used as a platform for remote epitaxy in creating free-standing oxide nanomembranes, opening an unprecedented way for functional oxide device applications[8,9].

However, characterizing graphene layers placed on $SrTiO_3$ substrates has been a formidable task. Raman spectroscopy is one of the standard non-destructive tools for inspecting the properties of graphene layers since Raman spectral features can tell us about the sample's defect densities and types[10], thermal properties[11], doping levels[12], and lattice strain[13,14]. For example, there are three well-known Raman scattering processes of graphene, the so-called G mode (i.e., the $E_{2g}$ zone-center scattering), D mode (i.e., the in-plane $A_{1g}$ zone-edge intervalley scattering), and 2D mode (i.e., an overtone of the D mode). Note that the G, D, and 2D modes are relevant to free-carrier doping, defects, and strain, respectively[15]. Observation of these peaks in Raman spectra provides a comprehensive understanding of lab-prepared graphene samples. Nevertheless, this approach has had limitations for graphene/$SrTiO_3$ samples because a strong multiphonon scattering of SrTiO$_3$ makes both the D mode and G mode peaks untraceable in conventional micro-Raman spectroscopic measurements, as shown in Fig. S1.

In this letter, we report that confocal Raman spectroscopy combined with simple spectral subtraction can be used effectively for characterizing graphene layers placed on SrTiO$_3$ single crystals. By aligning the confocal plane of the laser beam, we can obtain the Raman spectra of graphene layers and SrTiO$_3$ substrates, respectively, with a substantially reduced spectral intensity overlap. This approach of confocal Raman spectroscopy overcomes the drawback of conventional micro-Raman spectroscopic measurements and reveals inelastic light scattering peaks of graphene layers, enabling quantitative spectral analysis. For example, Raman spectra of two different graphene samples that are synthesized using silicon carbide (SiC) and germanium (Ge), respectively, and transferred to SrTiO$_3$ substrates indicate that they have distinct defect densities and types. Temperature-dependent Raman spectra reveal that graphene layers on SrTiO$_3$ experience strain from the substrate that is different from either silicon (SiO$_2$/Si) substrates or copper (Cu) foils. The outcome of this confocal Raman spectroscopic approach on graphene/SrTiO$_3$ provides indispensable information not only in understanding the material but also in the device application for remote epitaxy of functional oxides.

Figure 1 illustrates how the Raman spectrum of the graphene layer on SrTiO$_3$ is obtained using confocal Raman spectroscopy. We used a confocal micro-Raman spectrometer (JobinYvon LabRam HR800) with a 633-nm laser excitation to obtain inelastic light scattering spectra of our graphene/SrTiO$_3$ samples. A well-defined confocal plane along the sample's surface normal direction with a vertical accuracy of ~0.5 $\mu$m (Refs. 16 and 17) allows individual spectral measurements of the graphene layer and the SrTiO$_3$ substrate, respectively, as shown in the schematic diagram of Fig. 1. Note that both measurements are carried out on a single

graphene/SrTiO$_3$ sample without a need of additional SrTiO$_3$ reference measurement. We can distinguish the spectral features of graphene (red curve) and SrTiO$_3$ (blue curve) even though there are some common peaks near 1200 – 1700 cm$^{-1}$ between the two. The graphene-focused spectrum (red curve) also exhibits the multiphonon peaks of SrTiO$_3$ due to the confocal plane being thicker than the graphene layer, i.e., 0.5 $\mu$m > 0.35 nm. Nevertheless, by subtracting the SrTiO$_3$-focused spectrum (blue curve) from the graphene-focused spectrum (red curve), we can obtain the graphene-only spectrum (black curve), which is consistent with the previously reported Raman spectrum of graphene[18-20]. Note that the spectrum of Fig. 1 clearly shows weak features such as the D and D' modes, which are invisible in conventional micro-Raman spectroscopic measurements, as shown in Fig. S1.

**Table 1:** Raman peak intensity ratios for the two different graphene/SrTiO$_3$ samples synthesized using different graphene synthesis methods.

|  | $I_D/I_{D'}$ | $I_D/I_G$ | $I_{2D}/I_G$ |
|---|---|---|---|
| Graphene(SiC) | 2.7 | 0.7 | 2.4 |
| Graphene(Ge) | 4.5 | 3.4 | 0.3 |

We measured a few different graphene/SrTiO$_3$ samples to see if this approach is effective in examining their qualities since the Raman spectral features are correlated with the properties of defects in graphene. We used graphene layers that were respectively grown on SiC and Ge via silicon sublimation of the Si-terminated face of SiC and chemical vapor deposition on hydrogen-terminated Ge and transferred to the surface of SrTiO$_3$ substrates using a dry transfer method as reported in Refs. 8, 9, and 21. Figure 2 shows the graphene-only spectra of two different samples, i.e., graphene(SiC)/SrTiO$_3$ and graphene(Ge)/SrTiO$_3$. Note that both the D mode and D' mode (i.e., the in-plane $A_{1g}$ zone-edge intravalley scattering) peaks of the graphene(Ge)/SrTiO$_3$ sample are significantly higher than those of graphene(SiC)/SrTiO$_3$, implying that the former has larger defect densities than the latter. The intensity ratios between the D and D' modes ($I_D/I_{D'}$) are approximately 4.5 (for graphene(Ge)/SrTiO$_3$) and 2.7 (for graphene(SiC)/SrTiO$_3$), being smaller than 7 (See Table 1.), implying that both samples possess predominantly vacancy-type defects rather than $sp^3$-type defects, as discussed in Refs. 10 and 22. Nevertheless, the intensity of the 2D mode normalized by the G mode, i.e., $I_{2D}/I_G$, from the graphene(SiC)/SrTiO$_3$ sample is approximately eight times larger than that of the graphene(Ge)/SrTiO$_3$ sample, indicating that the former is higher quality graphene than the latter[10,22].

Temperature-dependent confocal Raman spectra show that the graphene layer on SrTiO$_3$ substrates exhibits a distinct temperature coefficient compared to other graphene layers transferred to SiO$_2$/Si substrates and grown on Cu foils. Figure S2 shows temperature-dependent Raman spectra of a graphene(SiC)/SrTiO$_3$ sample from room temperature down to 10 K using a custom-built optical cryostat. All Raman modes of the graphene are shifted to higher energies at low temperatures, which is qualitatively consistent with the previous reports of Refs. 11, 23, and 24. Figure 3(a) shows the temperature dependence of the G mode energies, whose slope ($\Delta\omega_G$) is approximately -0.049 cm$^{-1}$K$^{-1}$. Note that this slope value is in between those of graphene layers placed on SiO$_2$/Si, i.e., -0.016 cm$^{-1}$K$^{-1}$ (Ref. 11), and Cu, i.e., -0.101 cm$^{-1}$K$^{-1}$ (Ref. 23), as shown in Fig. 3(a). This observation makes sense because these substrate materials have different thermal expansion coefficients, i.e., $2.6\times10^{-6}$ K$^{-1}$ (SiO$_2$/Si) < $9.0\times10^{-6}$ K$^{-1}$ (SrTiO$_3$) < $16.5\times10^{-6}$ K$^{-1}$ (Cu)[25]. Figure 3(b) shows the temperature dependence of the D and 2D mode energies. Note that both the G mode (in Fig. 3(a)) and D mode (in Fig. 3(b)) are blue-shifted by 15 cm$^{-1}$ from room temperature to 10 K, whereas the 2D mode is shifted by 31 cm$^{-1}$. This unequal temperature-dependent peak shift is understood as the dominant biaxial tensile-strain on the graphene layer by SrTiO$_3$ substrates, which is estimated to be approximately 0.23% (Refs. 26 and 27).

In conclusion, the inelastic light scattering spectrum using confocal Raman spectroscopy provides useful information in understanding the graphene layers under various conditions. We obtained the information on defects and strain states for graphene/SrTiO$_3$ samples, which had been difficult to characterize due to the strong spectral overlap between the graphene and SrTiO$_3$. We suggest that this approach will shed light on the investigation of not only this graphene/SrTiO$_3$ system but also various 2D materials where their substrate's Raman modes interfere with those of the 2D materials.

**Acknowledgments**

We thank Chengye Dong, Joshua Robinson, Ji-Yun Moon, Jae-Hyun Lee, and Bernhard Keimer for providing samples, experimental help, and valuable discussions. The work at the University of Kentucky was supported by National Science Foundation Grant No. DMR-2104296. The team at MIT acknowledges support from the Air Force Research Laboratory (Award No. FA9550-19-S-0003). N.L.K. was partially supported by Southern Company. A.S. thanks Hian for the careful correction of this manuscript.

**Figure Captions**

**Fig 1:** Raman spectra of graphene(SiC)/SrTiO$_3$ when the confocal plane of the laser beam is focused on the graphene layer (red curve) and the SrTiO$_3$ substrate (blue curve), respectively. These two spectra are shifted for clarity. The spectrum of graphene (black curve) is obtained by subtracting the blue curve from the red curve. The inset shows a schematic diagram of the laser beam focus in confocal Raman spectroscopy. The spectrum of SrTiO$_3$ is obtained by placing the confocal plan 20 μm below the graphene layer.

**Fig 2:** Raman spectra of the graphene layers transferred from Ge (orange curve) and SiC (green curve) to SrTiO$_3$ single crystals.

**Fig 3:** Temperature-dependent peak-energy shifts of **(a)** the G mode and **(b)** the D and 2D modes of graphene/SrTiO$_3$ from room temperature to 10 K, i.e., $\Delta\omega \equiv \omega(T) - \omega(296\ \text{K})$. The peak energies are obtained by the Gaussian fit of each spectrum as shown in the inset. Temperature-dependent data of graphene layers placed on silicon substrates (dotted line, Ref. 11) and copper foils (dashed line, Ref. 23) are shown for comparison.

[26] Tao Jiang, Zuyuan Wang, Xiulin Ruan, and Yong Zhu, 2D Materials **6** (1), 015026 (2018).

[27] Charalampos Androulidakis, Emmanuel N. Koukaras, John Parthenios, George Kalosakas, Konstantinos Papagelis, and Costas Galiotis, Scientific Reports **5** (1), 18219 (2015).

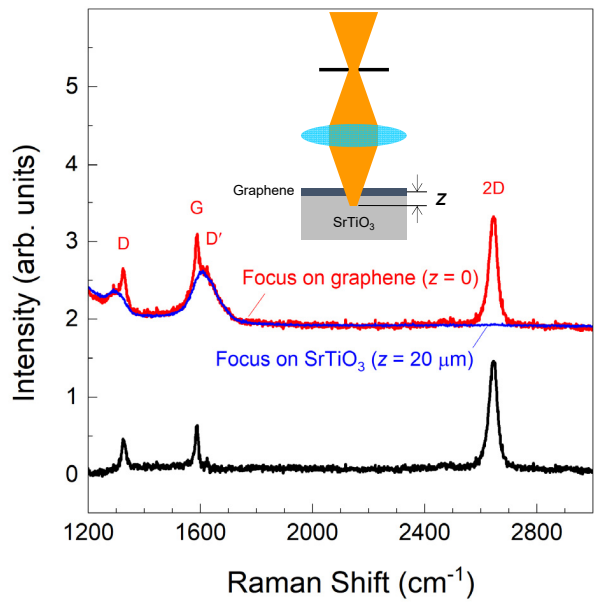

Figure 1

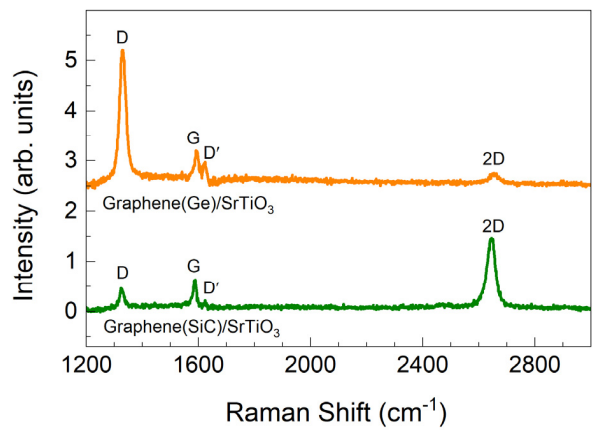

Figure 2

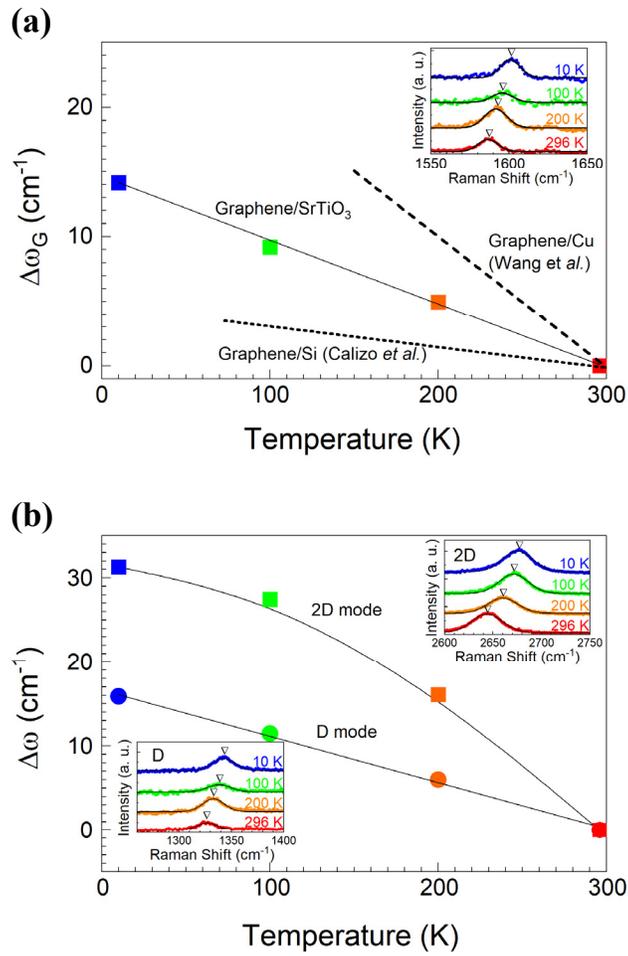

Figure 3

# Supplementary Material

**Investigations of Graphene on SrTiO$_3$ Single-Crystal using Confocal Raman Spectroscopy**

S. Shrestha[1], C. S. Chang[2,3], S. Lee[2], N. L. Kothalawala[4], D. Y. Kim[4], M. Minola[5], J. Kim[2], A. Seo[1]

[1]*Department of Physics and Astronomy, University of Kentucky, Lexington, KY 40506, USA*
[2]*Department of Mechanical Engineering, Massachusetts Institute of Technology, Cambridge, MA 02139, USA*
[3]*Research Laboratory of Electronics, Massachusetts Institute of Technology, Cambridge, MA 02139, USA*
[4]*Department of Chemistry, University of Kentucky, Lexington, KY 40506, USA*
[5]*Max-Planck-Institut für Festkörperforschung, D-70569 Stuttgart, Germany*

For comparison with our confocal Raman spectroscopic results, conventional micro-Raman spectroscopic measurements were carried out at 532 nm excitation using a Thermo-Scientific DXR micro-Raman spectrometer, as shown below.

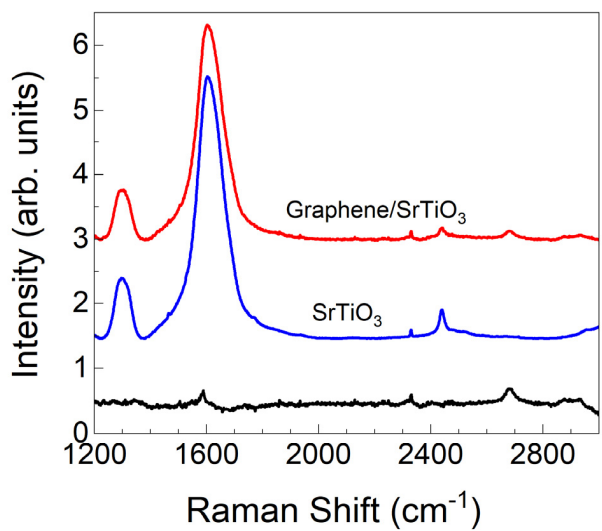

**Figure S1.** Conventional Raman spectra of graphene/SrTiO$_3$ (red curve) and bare SrTiO$_3$ (blue curve). The black curve is obtained by subtracting the bare SrTiO$_3$ spectrum from the graphene(SiC)/SrTiO$_3$ spectrum.

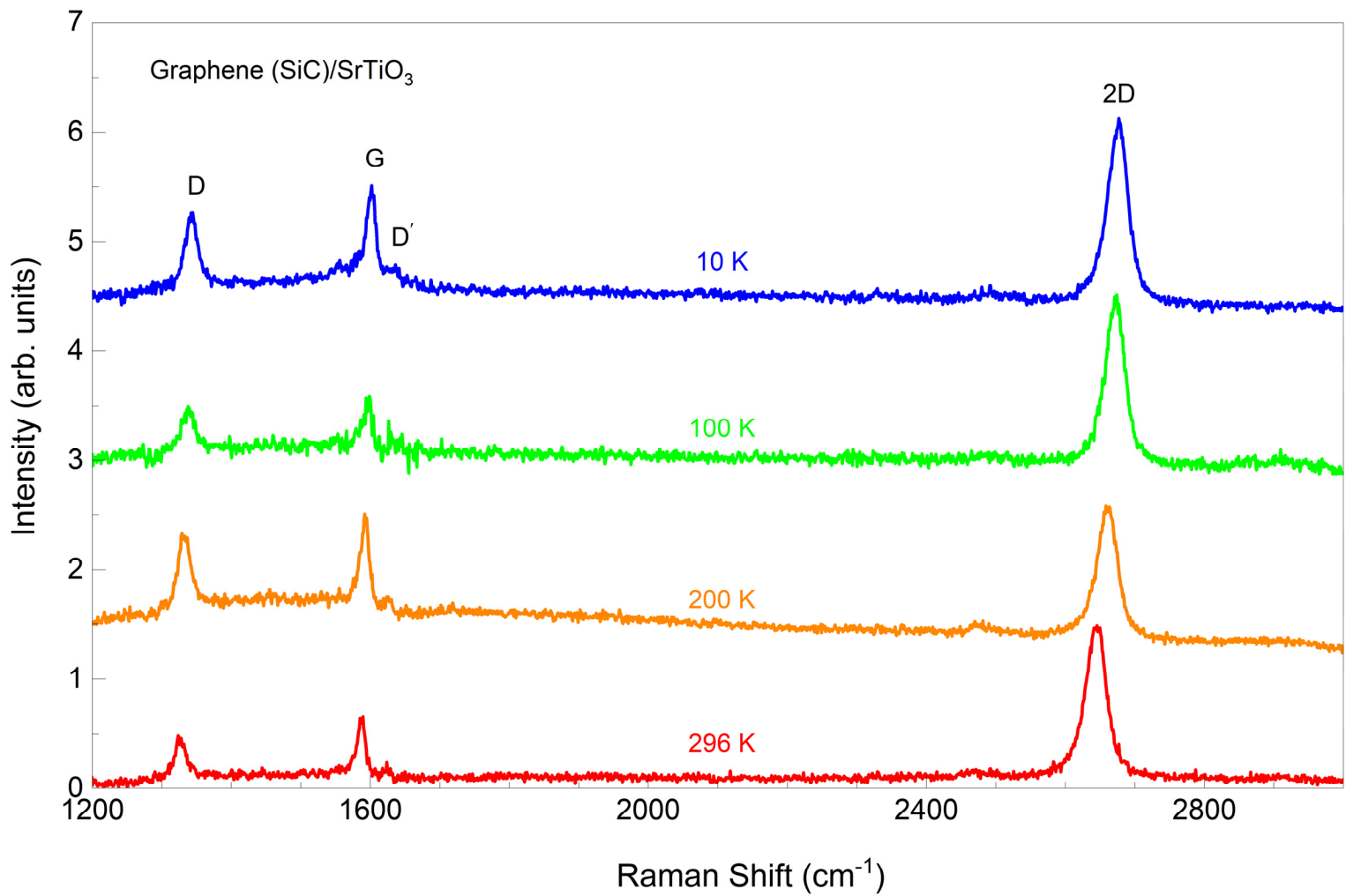

**Figure S2.** Temperature-dependence Raman spectra of graphene(SiC) obtained by confocal Raman spectroscopic measurements. Each temperature spectrum is shifted for clarity.